\documentclass[aps,pre,twocolumn,amsmath,amssymb,showpacs,amsfonts]{revtex4}
\usepackage{epsfig}
\usepackage{graphicx}
\usepackage{dcolumn}
\usepackage{bm}

\newcommand{\beq}{\begin{equation}}
\newcommand{\eeq}{\end{equation}}
\newcommand{\bea}{\begin{eqnarray}}
\newcommand{\eea}{\end{eqnarray}}
\def\half{{\scriptstyle{\frac{1}{2}}}}

\begin{document}

\title{The Incompressible Navier-Stokes Equations From Black Hole Membrane Dynamics}

\author{Christopher Eling$^1$}
\author{Itzhak Fouxon$^2$}
\author{Yaron Oz$^2$}

\affiliation{$^1$ Racah Institute of Physics,
  Hebrew University of Jerusalem,
  Jerusalem 91904, Israel}
  \affiliation{$^2$ Raymond and Beverly Sackler School of
Physics and Astronomy, Tel-Aviv University, Tel-Aviv 69978, Israel}

\date{\today}

\begin{abstract}

We consider the dynamics of a $d+1$ space-time dimensional membrane
defined by the event horizon of a black brane in $(d+2)$-dimensional
asymptotically Anti-de-Sitter space-time and show that it is
described by the $d-$dimensional incompressible Navier-Stokes
equations of non-relativistic fluids. The fluid velocity corresponds
to the normal to the horizon while the rate of change in the fluid
energy is equal to minus the rate of change in the horizon
cross-sectional area. The analysis is performed in the Membrane
Paradigm approach to black holes and it holds for a general
non-singular null hypersurface, provided a large scale hydrodynamic
limit exists. Thus we find, for instance, that the dynamics of the
Rindler acceleration horizon is also described by the incompressible
Navier-Stokes equations. The result resembles the relation between
the Burgers and KPZ equations and we discuss its implications.

\end{abstract}


\maketitle

The dynamics of fluids has remained an unsolved problem for
centuries. Unlocking its main features, chaos and turbulence, is
likely to provide an understanding of the principles and non-linear
dynamics of a large class of systems far from equilibrium. The
fundamental formulation of the problem of the nonlinear dynamics of
fluids is given by the incompressible Navier-Stokes (NS) equations
\cite{Landau,Frisch}
\begin{equation}
\partial_t v_i + v_j\partial_j v_i=-\partial_i P+\nu\partial_{jj} v_i + f_i \ , \label{NS}
\end{equation}
where $v_i(x, t), i=1,...,d,$ ($d \geq 2$) obeying $\partial_i v_i
=0$ is the velocity vector field,  $P(x,t)$ is the fluid pressure
divided by the density, $\nu$ is the (kinematic) viscosity and
$f_i(x,t)$ are the components of an externally applied force. The
equations can be studied mathematically in any space dimensionality
$d$, with two and three space dimensions having an experimental
realization. Among the central questions posed by the NS equations
is the existence of singularities in the solutions and the
statistics of the solutions in the limit of small $\nu$, both with
and without forcing.

The scarce progress achieved in understanding the NS equations
prompts one to look for other frameworks of viewing the dynamics
that might offer new insights. In this letter we explore a new
viewpoint that can lead to a development of geometrical methods to
study the NS equations. We represent the whole spatio-temporal
picture of the velocity field in terms of a structure of a null
hypersurface (whose normal vector is also a tangent vector),
embedded in a bulk space-time and evolving according to the General
Relativity field equations. Thus, the dynamics of the NS equations
is related to the dynamics of the geometry, as described by the
relativistic Einstein equations.

A particular case of a geometric representation of the NS equations
has been emerging recently based on the AdS/CFT correspondence
relating conformal field theories (CFTs) in a flat
$(d+1)$-dimensional space-time to gravity (string) theory on
asymptotically $(d+2)$-dimensional Anti de Sitter (AdS) space-time
\cite{Maldacena:1997re} (for a review see \cite{Aharony:1999ti}).
Here the spatio-temporal pattern of fluid motion is mapped onto a
black brane solution (a black hole with planar topology) in an
asymptotically AdS space-time. The map is based on the realization
that at large spatial and temporal scales the CFT dynamics reduces
to hydrodynamics, and the dual gravity description consist of long
wavelength, long time perturbations of the black brane solution
\cite{Bhattacharyya:2008jc}.

This correspondence has become relevant to the study of ordinary
fluids since it has been shown that hydrodynamics of any CFT,
although relativistic intrinsically, does admit a non-relativistic
(slow motion) limit given by the incompressible Navier-Stokes (NS)
equations describing flows of non-relativistic fluids
\cite{Fouxon:2008tb,Bhattacharyya:2008kq,Fouxon:2008ik}.  Thus,
applying the correspondence between the black brane solutions of
gravity and the flows of the CFT in the non-relativistic sector of
the latter, allows one to construct a geometrical description of the
NS equations.

The way the black brane horizon geometry encodes the boundary fluid
dynamics is reminiscent of the {\it Membrane Paradigm} in classical
general relativity, according to which any black hole has a
fictitious fluid living on its horizon \cite{membrane}. In the
AdS/CFT correspondence, the real fluid whose dynamics we wish to
study is at the boundary of the space-time. It is natural to ask to
what extent it can be identified under the duality map with the
membrane paradigm fluid (see e.g.
\cite{Kovtun:2003wp,Iqbal:2008by}).

We will consider this question in a rather general setup. Using the
Membrane Paradigm approach as developed in \cite{Dam} we will first
analyze the dynamics of a membrane defined by the event horizon of a
black brane in $(d+2)$-dimensional asymptotically AdS space-time,
showing that it is described by the incompressible NS equations of
non-relativistic fluids. The fluid velocity corresponds to the
normal to the deformed event horizon, while the rate of fluid energy
change is minus the rate of change in the horizon cross-sectional
area.

The analysis that we will perform also holds for any non-singular
null hypersurface when a large scale hydrodynamic limit exists. Thus
we will find, for instance, that the dynamics of the Rindler
acceleration horizon is also described by the incompressible
Navier-Stokes equations.

The connection between the horizon hypersurface dynamics and the NS
equation is analogous to the connection between the Burgers
\cite{Burgers} and the Kardar-Parisi-Zhang (KPZ) equations
\cite{Kardar:1986xt}. Our result shows that real turbulence may also
be seen as resulting from a physically natural surface dynamics.

In the following we will use the convention $8 \pi G=c=\hbar=k_B=1$.
We will consider a $(d+2)$-dimensional bulk space-time $M$ with
coordinates $X^A, A=0,...,d+1$ with a Lorentzian metric $g_{AB}$.
Let $H$ be a $(d+1)$-dimensional null hypersurface (notion akin to
horizon) characterized by the null normal vector $\bm n$ which
components $n^A$ obey \beq \bm n \cdot \bm n = g_{AB} n^A n^B = 0
\label{normal} \ . \eeq Note that this condition implies that for a null
hypersurface the normal vector is also a tangent vector. We define
the hypersurface in the bulk space-time by $x^{d+1} \equiv r =
const$, and denote the special coordinate system of \cite{Dam} as
$x^{\mu} = (t, x^i), i=1,...,d$. The coordinate $t$ parameterizes a
slicing of space-time by spatial hypersurfaces and $x^i$ are
coordinates on sections of the horizon with constant $t$. In this
coordinate system $n^r=0$ and one may choose the normalization
$n^t=1$. We identify the remaining components with the fluid
velocity in Eq.~(\ref{NS}) \beq n^i = v^i \ . \eeq

In the case of black branes in AdS, the event horizon of a static,
unperturbed, equilibrium solution is located in the bulk space-time
at $r=\pi T_0$, where $T_0$ corresponds to the Hawking temperature.
The horizon coordinates $(t,x^i)$ can be identified with time and
space Eddington-Finkelstein (EF) coordinates in the AdS boundary. We
consider slowly varying (long wavelength, long time) perturbations
of the static black brane solution, so we work to leading order in
an expansion in horizon derivatives. We also consider the slow
motion limit where $v^i$ is a small perturbation. Temporarily
restoring $c$, the horizon coordinates become $(ct, x^i)$ and the
non-relativistic slow motion limit corresponds to $v^i/c \ll 1$. In
what follows, in order to keep track of the different terms we
impose the scaling $\partial_t \sim \varepsilon^2, v^i \sim
\partial_i \sim \varepsilon$, where $\varepsilon$ is a small
parameter corresponding to $c^{-1}$ \footnote{Note, that this is not
the only conceivable slow motion, long distance scaling limit; one
could imagine $v^i$ and $\partial_i$ scaling differently from each
other. However, it turns out that our particular scaling is a
natural choice because it is a symmetry of the incompressible NS
equations \cite{Bhattacharyya:2008kq}.}. To $O(\varepsilon)$ the
$(d+2)$ black brane metric in EF coordinates is
\bea ds^2 = -r^2 f dt^2+ 2 dt dr + r^2 \sum^{d}_{i=1} dx^i dx_i  \nonumber \\
-\frac{2 \pi^4 T_0^4}{r^2} v_i dx^i dt - 2 v_i dx^i dr,
\label{bulkBB}\eea
where $f = 1 - \frac{\pi^4 T_0^4}{r^4}$.

The first fundamental form of the horizon is the space-time metric
restricted to it. In the horizon coordinate system it takes the
general form
\beq ds^2_H = h_{ij} (dx^i - v^i dt) (dx^j - v^j dt) \ , \eeq
where $h_{ij}$ is the metric on sections $S$ of the horizon $H$ at
constant $t$. In our case, we work to leading order in
$\varepsilon$, considering only the terms zeroth order and linear in
$v^i$. For the black brane, at zeroth order
\beq h^{(0)}_{ij} = (\pi T_0)^2 \delta_{ij} \ .
\label{hormetric}\eeq
The details of the subleading term, which is of order
$\varepsilon^2$, will not be needed for our analysis.

The second fundamental form of the horizon hypersurface is the
extrinsic curvature. In order to construct it consider the
space-time covariant derivative $\nabla_{A}$ projected into the
horizon surface and acting on the normal vector $n^A$. Since $\bm n
\cdot \bm n = 0$ (\ref{normal}) we have in our coordinates
\beq n_A \nabla_{\mu} n^A = 0 \ . \eeq
This implies that $\nabla_{\mu} n^A$ is tangent to the horizon and
can be expanded in the horizon basis $e^A_{\mu}$
\beq \nabla_{\mu} n^A = K_{\mu}^{\nu} e^A_{\nu} \ , \eeq
where $K_{\mu}^{\nu}$ acts as the extrinsic curvature of the
horizon. Together, the first and second fundamental forms provide a
complete description the embedding of the null hypersurface in the
bulk space-time.

Consider Lie transport of $h_{ij}$ along the null normal vector $\bm
n$, which is given by the Lie derivative ${\cal L}_{\bm n}$
\beq {\cal L}_{\bm n} h_{ij} = \half \partial_t h_{ij} + \half (\pi
T_0)^2 (D_i v_j + D_j v_i) \ , \eeq
where $D_i$ is the covariant derivative with respect to the metric
$h_{ij}$. Note that the $i,j$ indices are lowered and raised by the
horizon metric $h_{ij}$ in (\ref{hormetric}). The above expression
can be split into its trace part (the expansion $\theta$)  and trace
free part (the shear $\sigma_{ij}$)
\bea \theta &=& \half h^{ij} \partial_t h_{ij} + D_j v^j \ , \\
\sigma_{ij} &=& {\cal L}_{\bm n} h_{ij} - \theta h_{ij}/d \ . \eea
For the black brane we get to leading order the $O(\varepsilon^2)$
expressions
\bea \theta &=& \partial_i v_i \ , \\
\sigma_{ij} &=&  \half (\pi T_0)^2 \left(\partial_iv_j+\partial_jv_i
- 2\partial_k v_k \delta_{ij}/d \right) \label{shear} \ . \eea
Using $\bm n$ and $e^A_{i}$ as a tangent basis,  the components of
the horizon extrinsic curvature are generally
\bea K^{n}_{n} &=& \kappa(x) \ ,\\
K^{n}_{i} &=& \Omega_i \ , \\
K^{i}_j &=& \sigma^i_j + \theta \delta^i_j/d \ . \eea
$\kappa(x)$ is the surface gravity defined by $n^B \nabla_B n^A =
\kappa(x) n^A$. We parameterize it as \beq \kappa(x) = 2 \pi T_0(1 +
P(x) - v^2/2) \ , \label{kappa} \eeq where $P(x)$ and $v^2$ scale as
$\varepsilon^2$. We will identify $P(x)$ as the fluid pressure.
$\Omega_i$ is defined by \beq \Omega_i = m^A \nabla_i n_A \ , \eeq
where $m^A$ is an auxiliary null vector such that $m^A n_A = 1$.
Using (\ref{bulkBB}), for the black brane we find leading order \beq
\Omega_i = 2 \pi T_0 v_i \ . \eeq

We assume that the dynamics of the horizon geometry perturbations
are governed by the Einstein equations. The black brane is a
solution to the Einstein equations with negative cosmological
constant
\beq R_{AB} + (d+1) g_{AB} = T_{AB}^{matt} \ .\eeq
The Ricci tensor contraction $R_{AB} \ell^A e^B_\mu$ can be
expressed completely in terms of the horizon metric and extrinsic
curvature \cite{Dam}. Consider first the component along $\bm n$,
which is the contraction with $n^A n^B$.  This is the null focusing
equation, which after imposing the Einstein equation takes the form
\beq - n^A \nabla_A \theta + \kappa(x) \theta - \theta^2/d -
\sigma_{AB} \sigma^{AB}  = T_{AB}^{matt} n^A n^B \ .
\label{focusing} \eeq
There is no contribution from the cosmological constant term
proportional to the metric due to (\ref{normal}). In the black brane
case $T_{AB}^{matt} = 0$. Plugging our previous results for the
expansion, shear, and surface gravity into (\ref{focusing}), we find
at leading order the incompressibility condition
\beq
\partial_i v_i = 0 \ .
\eeq

Now consider the components of the Ricci tensor contraction
transverse to $\bm n$. By definition, there is again no contribution
from the cosmological constant term. From the remaining terms one
finds
\beq {\cal L}_{\bm n} \Omega_i + \partial_i \kappa(x) - D_j
\sigma^j_i + \frac{1}{d}
\partial_i \theta = -n^A e^B_i T^{matt}_{AB} \ , \label{horizonNS}\eeq
where
\beq {\cal L}_{\bm n} \Omega_i = (\partial_t + \theta)\Omega_i  +
v^j D_j \Omega_i + \Omega_j D_i v^j \ . \eeq
When $T_{AB}^{matt} = 0$ we find that the leading order terms are at
order $\varepsilon^3$ and give the NS equation (\ref{NS}) without a
force term and with a kinematic viscosity $\nu = (4\pi T_0)^{-1}$.
In cases where there is a non-zero matter stress tensor it will act
as a forcing term in the NS equations, the force being \beq f_i =
T_{AB}^{matt} n^B e^A_{i} \ . \eeq For instance, adding a dilaton
$\phi$ to gravity, results in \beq f_i = \nabla_A \phi \nabla_B \phi
n^B e^A_{i}/2 \ . \eeq

In the following we will consider the NS equations without a forcing
term. From the NS equations one can derive the energy balance
equation
\beq \int \half \partial_t v^2 d^d x = - \int \nu \partial_i v_j
\partial^i v^j d^d x  \ , \label{NSdiss}\eeq
that relates the rate of change of the fluid energy to minus the
energy dissipation per unit time due to fluid friction. To interpret
this equation in terms of the horizon geometry, consider the
focusing equation (\ref{focusing}) expanded to order $\varepsilon^4$
with $T^{matt}_{AB}=0$,
\beq 2\pi T_0 \theta = \sigma_{AB} \sigma^{AB}. \label{4thfocusing}
\eeq
The expansion of the horizon is defined as the fractional rate of
change in the cross-sectional area along the horizon generators \beq
\theta =  {\cal L}_{\bm n} \ln \sqrt{h} \ , \eeq where ${\cal
L}_{\bm n}$ is the Lie derivative and $h$ is the determinant of
$h_{ij}$. Using the incompressibility condition, (\ref{shear}) for
the shear, and (\ref{hormetric}) for the metric and then integrating
(\ref{4thfocusing}) over a horizon cross-section one has
\beq \partial_t A = \nu (\pi T_0)^d \int
\partial_i v_j \partial^i v^j ~ d^{d} x \ ,
\label{eqone} \eeq
where $A$ is the total horizon area. Imposing the energy balance law
(\ref{NSdiss}), we find that
\beq \partial_t \left(A/A_0\right) = - \int \partial_t  v^2/2 ~
d^{d} x, \label{areadiss}\eeq
where $A_0$ is the zeroth order area density $(\pi T_0)^d$. Thus, as
the kinetic energy of the fluid on the boundary decreases in time
due to viscous dissipation, the horizon area grows. This is
consistent with the classical area increase theorem of General
Relativity \cite{Hawking:1971tu}.

The derivation of the NS equations required knowledge of the horizon
embedding and employed a local analysis near this horizon. Unlike
\cite{Bhattacharyya:2008kq}, there was no need to know the
asymptotic structure of the full bulk space-time. Therefore the
results will apply to perturbations of a stationary non-singular
null hypersurface, as long as there is a separation between the
characteristic scale $L$ of the perturbations and some intrinsic
microscopic scale. The non-singularity requirement was used when
contracting the Einstein equations in order to obtain the membrane
equations. To see why a separation of scales is crucial, consider
for example black holes in asymptotically flat spaces with horizons
of a spherical topology (e.g. Schwarzschild). By dimensional
analysis the correlation length of a fluid will scale as $l_c \sim
T_0^{-1}$, where $T_0$ is the Hawking temperature. In the
asymptotically flat cases $T_0^{-1} \sim r_0$, where $r_0$ the
horizon radius. Since the horizon is now also compact, the $L$ can
be no greater than $\sim r_0$. Thus the dimensionless Knudsen number
$Kn\equiv l_{c}/L$ is of order unity, implying that the derivative
expansion used above is not valid and that hydrodynamics is not the
appropriate effective description. Moreover, in the general NS-like
equation (\ref{horizonNS}) found in the Membrane Paradigm approach,
the term $\partial_i \theta$ does not vanish and leads to the
assignment of an unphysical negative ``bulk viscosity"
\cite{membrane,Dam}.

To show that our analysis is applicable to cases other than the
black branes in AdS, consider the example of Rindler space
associated with accelerated observers in $(d+2)$-dimensional flat
Minkowski space-time \beq ds^2 = -\kappa^2 \xi^2 d\tau^2+ d\xi^2 +
\sum^{d}_{i=1} dx^i dx_i \ , \eeq where $\kappa$ is a constant. To
the uniformly accelerated observer with worldline $\xi = const.$,
the surface $\xi=0$ is a causal horizon, with intrinsic metric
$h_{ij} = \delta_{ij}$, that prevents him from an access to the
entire space-time. The constant $\kappa$ can be identified with a
temperature  in the following way. Unruh \cite{Unruh:1976db} showed
that accelerated observers feel the quantum vacuum to be a thermal
state at temperature $T=a/2\pi$, where $a$ the observer's proper
acceleration. This Tolman local temperature can be expressed as
$T=\kappa/2\pi \chi$, where $\chi= \sqrt{-g_{\tau \tau}} = \kappa
\xi$ is the redshift factor. We define $\kappa = 2\pi T_0$ as the
location independent temperature of the system.

To perturb this horizon we allow, as in the black brane example, a
slowly varying ($L \gg T_0^{-1}$) fluid velocity $n^i=v^i$ of $\sim
\varepsilon$. In EF coordinates the Rindler metric to
$O(\varepsilon)$ is
\bea ds^2 = -2\pi T_0 r dt^2 + dt dr + \sum^{d}_{i=1} dx^i dx_i \nonumber \\
- v_i dx^i dr - 4\pi T_0 r dx^i dt. \eea
We assume the perturbations are governed by the vacuum Einstein
equations with zero cosmological constant, $R_{AB} = 0$.
Parameterizing $\kappa(x)$ as in (\ref{kappa}) and applying the
membrane analysis to the Rindler horizon ($r=0$) also shows that its
dynamics is determined by the incompressible NS equations with
kinematic viscosity $(4\pi T_0)^{-1}$. In this case though the
result cannot be understood as mirroring the hydrodynamics of a
field theory fluid living on an asymptotic boundary of space-time.
Here the fluid system could be interpreted as the vacuum thermal
state (thermal gas of Rindler particles) perceived by accelerated
observers, with the horizon acting as the holographic screen
\cite{Eling:2008af}.

Consider now some of the implications of our results. First, exact
solutions of the NS equations such as vortices and others
\cite{Saffmann,Landau} are mapped immediately into geometrical
horizons. Second, finite time singularities, where fluid velocity
gradients diverge, are mapped into the naked curvature singularities
in the gravity description. Thus,  the extremely important problem
of singularities in the NS equations in three space dimensions
appears linked to a cosmic censorship principle in gravity, a fact
that may be of use in the future research of this long-standing
problem. Third, in the case of turbulence, while an instantaneous
picture of a flow is analytically unavailable, the spatially
averaged correlators seem to have a well-defined analytic form.
There is experimental and numerical evidence that in the ``inertial"
range of distance scales $r$ much larger than characteristic length
of viscous dissipation $l$ but much smaller than flow excitation
scale $L$, the flows exhibit a universal behavior, e.g. the
space-averaged equal-time correlators of velocity differences in the
inertial range are characterized by critical exponents. For
instance, the longitudinal n-point functions scale as
\begin{equation}
S_n(r) \equiv \langle \left[({\bf v}({\bf r})-{\bf v}(0))\cdot {\bf
r}/r\right]^n\rangle \sim r^{\xi_n} \ . \label{expon}
\end{equation}
The 1941 exact scaling result of Kolmogorov ${\xi}_3=1$ agrees well
with the experimental data. In the geometrical picture, $S_n(r)$
correspond to the space-averaged equal-time correlators of
differences of normals to the horizon. The fluid picture implies
that in the evolution of the horizon surface there is a range where
its area is preserved and it exhibits multi-fractal properties
characterized by the exponents $\xi_n$.

Multi-fractality of the horizon surface may provide a dynamical
basis for the so-called multifractal model of turbulence. This
model, which appears to be the most successful phenomenological
model of turbulence, expresses the anomalous exponents of turbulence
in terms of the spectrum of fractal dimensions \cite{Frisch,EFO}.
The model has been proposed originally on the basis of presumed
singularities of the dynamical equations \cite{FrischParisi}.
However, it has been recognized that this assumption may not be
correct \cite{Frisch,FrischPrivate}. This leaves the open question
of whether there is a singular structure underlying turbulence.
% and the singularities of what these are
%(with dissipation being a most natural candidate for a singular measure).
A possible natural answer within the presented geometrical picture
is that the horizon itself exhibits a multi-fractal structure in a
wide range of scales. For instance, within the approximation
corresponding to the Euler equation in hydrodynamics (infinite
$T_0$), the  dynamics may make the horizon surface indefinitely
rugged, while preserving its area. This could correspond to the
assumption of singularities of the Euler equation. Thus, a central
question is the horizon surface structure,
%does become more rugged or not, which means
which requires an understanding of the general features of the
surface fracturing. The geometrical picture is likely to be a useful
tool also in the  study of  the relation between the velocity
scaling in space and time, where in the latter one follows the
Lagrangian trajectory of a fluid particle in a turbulent flow and
considers moments of velocity difference at time $t$ \cite{Landau}.

\vskip 0.5cm

The work of I.F. and Y.O. is supported in part by the Israeli
Science Foundation center of excellence, by the Deutsch-Israelische
Projektkooperation (DIP), by the US-Israel Binational Science
Foundation (BSF), and by the German-Israeli Foundation (GIF). C.E.
is supported by the Lady Davis Foundation at Hebrew University and
by grant 694/04 of the Israel Science Foundation, established by the
Israel Academy of Sciences and Humanities.

\end{document}